\documentclass[11pt]{article}

\usepackage{cite}
\usepackage{amssymb}
\usepackage{amsmath}
\usepackage{bm}

\usepackage{graphicx}
\usepackage{epstopdf}
\def\~{\tilde}

\def\be{\begin{equation}}
\def\ee{\end{equation}}

\title{Quantum Mechanics: Harbinger of a Non-Commutative Probability Theory?}
\author{B. J. Hiley\footnote{E-mail address b.hiley@bbk.ac.uk.}.}
\date{\small {TPRU, Birkbeck, University of London, Malet Street, \\London WC1E 7HX.} }

\begin{document}
\maketitle

\begin{abstract}
In this paper we discuss the relevance of the algebraic approach to quantum phenomena first introduced by von Neumann before he confessed to Birkoff that he no longer believed in Hilbert space.  This approach is more general and allows us to see the structure of quantum processes in terms of non-commutative probability theory, a non-Boolean structure of the implicate order which contains Boolean sub-structures which accommodates the explicate classical world.  We move away from mechanical `waves' and `particles' and take as basic what Bohm called a {\em structure process}.  This enables us to learn new lessons that can have a wider application in the way we think of structures in language and thought itself.

\end{abstract}

\section{Introduction}

As Murry Gell-Mann \cite{mgell} once wrote:-

\begin{quote}
\small{Quantum mechanics, that mysterious, confusing discipline, which none of us really understands but which we know how to use.  It works perfectly, as far as we can tell, in describing physical reality, but it is a `counter-intuitive' discipline, as social scientists would say.  Quantum mechanics is not a theory, but rather a framework within which we believe a correct theory must fit}. \end{quote}
The professional physicist still finds explaining exactly what the quantum formalism is telling us about Nature very difficult.  We know that it is something to do with non-commutativity because the commutative algebra of functions used in classical physics is replaced by a non-commuting algebra of operators where the operators become `observables' while their eigenvalues correspond to the values found in experiments.  The formalism works, but what does it all mean?  Do the problems arise simply because of the small scale nature of the phenomena, or because the spectacular behaviour of matter only occurs at low temperature, with no general consequences for the way think about the macroscopic world in general?  Or is it pointing to something much more general which reaches down into the very being of our lives, providing a different paradigm that affects the way we think in general?

Let me begin with a personal difficulty.  I have aways been puzzled by the contrast between the way quantum theory was originally introduced and the way we worry over it today.  Quantum theory was introduced to explain two main phenomena, the stability of matter at room temperatures and the frequency spectrum of radiation coming from very hot objects like the sun.  Classical mechanics provides us with no explanation of the stability of an atom, the stability of a molecule, the stability of a biological cell, the stability of a solid crystal, the stability of my desk and so on.  We need quantum mechanics to provide the explanation.  We also need quantum mechanics to explain the radiation black body radiation.  Let me repeat, we needed quantum theory to explain the stability of large scale matter at room temperature and to explain effects emanating from very hot bodies.  Contrast this with what we worry about now.  We worry about the fragility of the coherence of the quantum state, we worry about schizophrenic cats, about the collapse of the wave function and we worry about the implications of quantum non-locality \cite{anton}.

These are the details we physicists are concerned with.  What we do not dispute is the novelty of the conceptual and mathematical form of the ideas that are involved.  Perhaps the most radical notion is that we must give up reductionism with its view that ultimately the world must be analysed into elementary parts and the relations between these parts define what we perceive to be the world around us.  To emphasise this failure of reductionism
consider our latest attempts to find the ultimate constituents of the nucleon.  Instead of finding simplicity, we find is a sea of activity which we analyse in terms of a multitude of partons comprising valence quarks, quark-antiquark pairs, gluons and perhaps even more \cite{Rob}.  There is no `ultimon' in which we pin the solidity of the macroscopic world.  

However we should not be surprised.  Bohr talks about the ``impossibility of any sharp separation between the behaviour of atomic objects and the interaction with the measuring instruments which serve to define the conditions under which the phenomena appear." \cite{nb61}.  In other words there exists a kind of wholeness in which ``we are not dealing with an arbitrary renunciation of a more detailed analysis of atomic phenomena, but with a recognition that such an analysis is {\em in principle} excluded." \cite{nb61a}.

Primas \cite{hp94} goes further 
\begin{quote}
\small{According to quantum mechanics the world is a whole, {\em a whole which cannot be made out of parts}.  If ones agrees that  quantum mechanics is a serious theory of matter then one cannot adopt the classical picture of physical reality with its traditional metaphysical presuppositions. In particular, the non-separability and non-locality of the material world and holistic features are not compatible with the ontology usually adopted in classical physics.}
\end{quote}
So where do we start?  Fortunately at the level of atoms and molecules we do have what appear to be autonomous objects, but these objects have both classical and quantum properties. Macroscopic molecules behave like classical objects but they combine through quantum processes, where, then, is the description that has both classical and quantum features?  

If we examine the mathematical structures we have, it appears as if classical mechanics is totally different from the mathematical description commonly used in quantum mechanics.  The variables used in classical mechanics, state functions of the position and momentum, $f(x, p)$,  have a product rule that is commutative
\begin{eqnarray}
f(x,p)\cdot g(x,p) -g(x,p)\cdot f(x,p) = 0	\label{eq:comm}
\end{eqnarray}

Whereas in quantum mechanics, $x$ and $p$ are replaced by operators that do not commute.  Similarly state functions, the density matrices, $\rho$, used in the algebraic form of quantum mechanics,  do not commute either.  Here I am using density matrices to describe the state of a system in contrast to the usual wave function because it is a more general approach which reduces to the usual approach only in a special case, namely, when $\rho$ is of rank one and idempotent, $\rho^2=\rho$.  Then we can write 
\begin{eqnarray*}
\rho=\psi^*\psi
\end{eqnarray*}
where $\psi$ is the wave function that causes us so many interpretational difficulties. What is important to notice here is the states themselves need not commute , namely,
\begin{eqnarray*}
\rho_1\rho_2-\rho_2\rho_1\ne 0
\end{eqnarray*}
This should be contrasted with the behaviour of classical state functions, equation (\ref{eq:comm}).
We immediately see that a fundamental mathematical difference between classical and quantum mechanics  lies in difference between the commutative and non-commutative structures.  We can already get a hint of how to combine these two aspects into a single structure if we ask,  ``Where do we see commutativity and non-commutativity occurring regularly in our everyday experiences?"  Not in the relations of things, but in the order of {\em action}.  For example, I cannot walk through a door without opening it first. The order of action is vital. For those of a philosophical turn of mind, it is Heraclitus not Democritus,  it is Schelling and Fichte, not Kant that provide the clues.  As Fichte \cite{jf94} writes
\begin{quote}
The question is whether philosophy should begin with the fact or an act (i.e., with a pure activity that presupposes no object, but instead, produces its own object, therefore with an acting that immediately becomes a deed).
\end{quote}

In this paper I want to use the algebraic formalism, because it is through this formalism that the real novelty of the quantum ideas and the connection with process come through showing us that there is a radically new way of looking at all aspects of life that leads us to abandon the classical paradigm and to replace it by a richer paradigm in which {\em structure process} is basic \cite{db65}.  

\section{Structure Process and Algebraic Order.}

When I started to work with David Bohm in the `60s he was thinking of how relativity and quantum theory could be brought together in a new way.  To avoid the difficulties of a rigid object presents to relativity, Bohm introduced the notion of a structure process in which a set of discrete ``space-like" elements undergo discrete or continuous changes as they move and unfold in a {\em process} of development.  He argued that such a notion implies that the structural process as a whole, with its set of manifold relationships of partial order of discrete elements, is logically and existentially prior to the notion of a continuous space-time, in the sense that the latter is an abstraction from the former,  representing a kind of approximate `map' of the overall structure process.  Thus the particles and fields, and indeed space-time itself were to be abstracted from this deeper process.  His discussions were conceptual and philosophical.  The question he left unanswered was how these notions could be developed into a coherent mathematical structure.

While I was thinking about these problems, I happened to come across two significant discussions.  The first was an essay by Hankins \cite{th76} who was reviewing Hamilton's work in which he introduced a notion,   ``the algebra of pure time".   Hamilton thought that ``in algebra the relations which we first consider and compare, are relations between successive changing thing or thought".  He then goes on ``Relations between successive thoughts thus viewed as successes states of one more general and changing thought, are the primary relations of algebra".  Note he uses `thought' not material process.  This raises the interesting notion that algebra is not only about material process as the physicist will believe; it has a much more general function, describing the order, not only of material relations, but also the order of thought.  Thought is not subject to the order of space-time. There is no notion of locality. Could this algebraic notion of order take us beyond the order of space-time revealing new relations of the type we see in entangled states?

Then there was Grasmann's {\em Ausdehnungslehre} \cite{gg84} that had a profound influence on Clifford's development of his algebra.  It is this algebra that I have found extremely fascinating and which forms a basis of my recent work on what I call {\em Bohm's non-commutative dynamics} \cite{bh13}\footnote{ Unfortunately I have to make it clear that the spirit of the view Bohm and I were developing together had little in common with the proponents of the subject now called ``Bohmian mechanics".}.  Grassmann introduced the notion of an {\em extensive} to carry the notion of a {\em continuous becoming}.  We all experience one thought transforming into another, new thought.  Is the new thought separate from the old thought?  No. The old thought contains the potentiality of the new thought, while the new thought contains a trace of the old thought. Symbolically this is written as $[T_1,T_2]$ then succession can be captured through a groupoid multiplication rule
\begin{eqnarray}
[T_1,T_i]\circ[T_j,T_3]=[T_1,T_3];	\quad \quad \mbox{only when}\; i=j. 	\label{eq:multi}
\end{eqnarray}
As I have shown elsewhere \cite{bh11}, encapsulated in this idea is the notion of unfolding that is central to the notion of enfolding and enfolding that leads directly to the Heisenberg equation of motion.  This is one of the equations that form the basis of the time development of quantum processes that we use implicitly throughout this paper.  

Let us first start by explain how these ideas lead us to Clifford algebras  \cite{bh11}.  Clifford \cite{wc87}, exploiting the ideas of Grassmann and Hamilton, introduced a multiplication rule, which he called {\em polar} multiplication, and which we now call {\em Clifford} multiplication.  This follows from equation (\ref{eq:multi}) together with
\begin{eqnarray*}
[T_1,T_2]=-[T_2,T_1]
\end{eqnarray*}
 Now one can easily show that the following rule is satisfied 
\begin{eqnarray*}
[T_1,T_2]\circ[T_2,T_3]+[T_2,T_3]\circ[T_1,T_2]=0
\end{eqnarray*}
showing that the product in anti-commutative.   Now let us consider the special case in which the thought turns into itself, that is $[T_1,T_1]$, the {\em idempotent} which formally satisfies
\begin{eqnarray*}
[T_1,T_1]\circ[T_1,T_1]=[T_1,T_1]
\end{eqnarray*}
This shows that the thought is not static, but keeps on turning into itself.  To show exactly how this structure produces the formal orthogonal Clifford algebra requires a little extra work which we will not need  in this paper so we  will simply refer to the original work of Clifford \cite{wc87} or to Hiley \cite{bh11} for a more extensive discussion of the ideas introduced here. 

However one thing that I will mention here to complete the background is to explain where Hamilton fits in.  Hamilton was interested in generalising the complex numbers, which would involve seeing how three mutually perpendicular two dimensional Argand planes can be fitted together into a three dimensional space.  Recall that the complex number $i$ can be regarded as a rotation through $90^0$ in, say, the $x-y$ plane.  How then do we combine this rotation with a $90^0$, rotation, say $j$, in the $x-z$ plain, and a $90^0$, say $k$, in the $y-z$ plain?  Clearly $i^2=j^2=k^2=-1$.  Notice we have introduced three separate, but related `square roots of $-1$'.  What Clifford showed was that if you take $[T_0,T_1]$ to be a movement along the $x$-axis and $[T_0,T_2]$ to be a movement along the $y$-axis, then $[T_1,T_2]$ is a movement (i.e., rotation) taking $T_1$ into $T_2$.  Clearly if you apply $[T_1,T_2]$ again you get $[T_1,T_2]^2$ which Clifford took to be $-1$, so that Hamilton's quaternions became a special case of a Clifford algebra.

\section{Where does Quantum Theory fit in?}

All of this mathematical structure was developed when classical mechanics was the only mechanics known.  Imagine the surprise when nature threw up spin and the Pauli $\sigma$ algebra and then Dirac  showed that a relativistic generalisation required the relativistic electron depended on a set of anti-commuting $\gamma$-matrices.   Both of these structures are examples arising in the tower of orthogonal Clifford algebras.

In an orthogonal Clifford algebra, ${\cal C}$, the rotations emerge as inner automorphisms defined by
\begin{eqnarray*}
A'=RAR^{-1},\quad\quad \forall A\in {\cal C}.
\end{eqnarray*}
where $R$ is a set of invertible elements in ${\cal C}$.  The multiplicative group, $G$ of invertible elements $R$ is called the Clifford group, which in the physics community is known as the spin group.  The Clifford group gives us direct access to the double cover of the usual rotation group and the spinor comes ``for free" as an element of a suitably chosen minimal left ideal.  It was through a detailed study of orthogonal Clifford algebras that Hiley and Callaghan \cite{bhbc10} extended the Bohm approach to relativistic particles with spin.

Now I want to draw your attention to another algebra which appeared in a classic paper by von Neumann \cite{jvn32}.  Again here we are not specifically concerned with quantum processes, but we arrive at an algebra that plays a central role in quantum mechanics.  Let us begin by considering the translations in an $(x,p)$ symplectic (phase) space.  We can write these translations as
\begin{eqnarray*}
\widehat U(\alpha)=\exp(i\alpha \widehat P);\quad \widehat V(\beta)=\exp(i\beta \widehat X)
\end{eqnarray*}
If the generator $\widehat P$ and $\widehat X$ are defined by the relations
\begin{eqnarray*}
\widehat U(\alpha) f(\widehat X)\widehat U(\alpha)^{-1}=f(\widehat X+\alpha); \quad \widehat V(\beta)g(\widehat P)V(\beta)^{-1}=g(\widehat P+ \beta)
\end{eqnarray*}
then $(\widehat X,\widehat P)$ must satisfy the relation $[\widehat X,\widehat P]=i$  where $[\cdot,\cdot]$ is the usual commutator\footnote{Of course position and momentum have different dimensions so we choose $x\leftrightarrow \widehat X$ and $p\leftrightarrow \epsilon \widehat P$. Note that we are not appealing to anything quantum mechanical at this stage. It is only in quantum mechanics that we write $\epsilon=1/\hbar$.}. 
We follow von Neumann and write
\begin{eqnarray*}
\widehat S(\alpha,\beta)=\exp i(\alpha \widehat P +\beta \widehat X)
\end{eqnarray*}
Here $\widehat S(\alpha,\beta)$ is the generator of the Heisenberg group acting in the symplectic space. 

 It is often believed that the Heisenberg algebra is a sign that we have entered the quantum domain, but this is not true.  Most of the exploration of the properties of this group are by people working in radar, which was, of course, designed to locate the position and speed of aircraft, hardly a quantum phenomena! 

If we interpret $(\widehat X,\widehat P)$ as the Hermitian operators used in the Hilbert space approach to quantum mechanics, then we see that the parameters $(\alpha, \beta)$ define a dual structure.  This dual structure contains all the information contained in the Hilbert space formalism, but in a novel way.  von Neumann shows there is a $1-1$ correspondence between the Hilbert space formalism and the functions $a(\alpha, \beta)$ through the relation
\begin{eqnarray}
\widehat A=\int\int a(\alpha, \beta)\widehat S(\alpha,\beta)d\alpha d\beta.	\label{eq:idty}
\end{eqnarray}
To obtain expectation values that agree with those formed in standard quantum mechanics, we introduce an element $\rho_s$, the density matrix and form 
\begin{eqnarray*}
f_\rho(\alpha, \beta)=Tr[\rho_s\widehat S(\alpha,\beta)]
\end{eqnarray*}
so that
\begin{eqnarray*}
\langle \widehat A\rangle=\int\int a(\alpha,\beta)f_\rho(\alpha,\beta)d\alpha d\beta.
\end{eqnarray*}
The form of this equation suggests that in the space defined by $(\alpha,\beta)$,  expectation values can be found in the same way as they are found in standard commutative statistics, however in this case the variables $(\alpha, \beta )$ are non-commutative, as we will soon see, so we have a generalisation of ordinary statistics.  This generalisation was suggested first by Moyal \cite{jm49} who further brought out the physical meaning of the approach by identifying $\alpha$ with $x$ and $\beta$ with $p$, so that we have a non-commutative phase space. In more general terms, a non-commutative symplectic space.  It is important to note that there exists  a commutative sub-space which contains classical mechanics.

I have called the space spanned by $(\alpha, \beta)$ non-commutative, but in what sense?  We have seen in equation (\ref{eq:idty}) there is a relation between $\widehat A\leftrightarrow a(\alpha,\beta)$ so if the operators do not commute, this must be reflected in the product $a(\alpha,\beta)\star b(\alpha,\beta)$.  Indeed von Neumann showed that it was necessary to introduce a new product
\begin{eqnarray*}
a(\alpha,\beta)\star b(\alpha,\beta)=\int\int e^{2i(\gamma\beta-\delta\alpha)}
a(\gamma-\alpha,\delta-\beta)b(\alpha,\beta)d\alpha d\beta
\end{eqnarray*}
Although this definition was introduced by von Neumann, it is now called a Moyal product because Moyal derived it in a more suitable form, namely,
\begin{eqnarray*}
a(x,p)\star b(x,p)=a(x,p)\exp[i\hbar(\overleftarrow\partial_x\overrightarrow\partial_p - \overleftarrow\partial_x\overrightarrow\partial_p)/2]b(x,p)
\end{eqnarray*}
It is easy to show that this product gives
\begin{eqnarray*}
x\star p-p\star x=i\hbar
\end{eqnarray*}
Since we have a non-commuting product, we can form two brackets; the Moyal bracket
\begin{eqnarray*}
\{a,b\}_{MB}=\frac{a\star b-b\star a}{i\hbar}
\end{eqnarray*}
This corresponds to the commutator bracket in standard quantum mechanics, and the Baker bracket
\begin{eqnarray*}
\{a,b\}_{BB}=\frac{a\star b+b\star a}{2}
\end{eqnarray*}
which corresponds to an anti-commutator.  

The interesting result is that in the limit $O(\hbar^2)$, we find the Moyal bracket becomes the Poisson bracket of classical mechanics
\begin{eqnarray*}
\{a,b\}_{MB}=\{a,b\}_{PB}=O(\hbar^2)=[\partial_x a\partial_p b-\partial_p a\partial_xb]
\end{eqnarray*}
While the Backer bracket becomes a commutative product
\begin{eqnarray*}
\{a,b\}_{BB}=ab+O(\hbar^2)
\end{eqnarray*}
Thus we find classical mechanics appearing as a sub-algebra in the non-commutative symplectic algebra.

Just to confuse matters, the algebra that we have outlined above goes under several different names.  Sometimes it is called the Weyl algebra \cite{gf89}, sometimes it is called the Moyal algebra \cite{cz05}, but, as we have shown, has it roots in the algebra introduced by von Neumann \cite{jvn32} in 1931.  I prefer to follow Crumeyrolle \cite{ac90} and call it the symplectic Clifford algebra because of its close relation with the orthogonal Clifford algebra. 
It is not appropriate to go into the details of the symplectic Clifford algebra in this paper.  We will simply point out a couple of features that I hope will stimulate some interest in the structure. 

Firstly the orthogonal Clifford algebra can be generated by the fermionic (Grassmann) creation and annihilation operators while the symplectic Clifford can be generated by the bosonic creation and annihilations operators.  Secondly the orthogonal Clifford group is the spin group so that spin occurs naturally in the algebra.  The symplectic Clifford group generates the double covering group of the symplectic group, the metaplectic group.  It `lives' in the covering space and accounts for phase properties like the Gouy  effect \cite{lg90, as86} and the Aharonov-Bohm  effect \cite{ab59}. 

The final point I would like to make is that both these algebras are geometric algebras in the sense that they describe deeper properties of the geometry of space-time that do not, {\em a priori}, depend on quantum theory.  Rather quantum phenomena exploit these deeper properties of space-time.  As these properties are global aspects of the underlying classical space-time, it should not be surprising to find non-local effects because there is no way of describing these global effects using only local Lie algebraic structures.

\section{Non-commutative Probability}

I am  arguing here that these global structures are not merely properties of the material world.  They have ramifications for all forms of activity, including the organising  orders in thought.  Recall that it was Hamilton and Grassmann thinking about the order of thought that led them to take algebraic structures seriously  in the first place.  The application of algebras has been very successful in the material word and it is only recently that people have been trying to apply the ideas in other areas releasing the formalism from  the shackles of the quantum theory.

What has held people back from exploiting quantum algebras has been the thinking that the only way to deal with quantum-like phenomena is through the Hilbert space formalism with its total dependence on the wave function.  The algebraic approach to quantum phenomena, while capturing all aspects of the theory that have already been explored, is a more general formalism and requires a different mind-set from that used in the Hilbert space formalism.  No one has worked harder to promote the algebraic approach than theoretical chemist, Hans Primas \cite{hp09,hp94}.  The general concept that he emphasises is that no matter how we mathematically analyse quantum phenomena, its essential feature is based on a non-commutative structures which he regards as needing a non-Boolean logic.  The classical world is Boolean and we have yet to understand the radically different attitudes needed to comprehend a non-Boolean way of thinking.

The two algebras that we have introduced in this paper are both non-Boolean   and  are specific examples of what are called
 von Neumann algebras.  Their  apparent very different  structure is because   the orthogonal Clifford is a type I  von Neumann algebra, while the symplectic Clifford is of type II.  A discussion of a type I algebra, based on the symplectic structure has been carried out in Hiley and Monk \cite{bhnm93} and in Bohm, Davies and Hiley \cite{dbpdbh}. I mention these technicalities because the difference in appearance of the two Clifford algebras might cause some uncertainty in the general discussion. In practice the precise differences are not important for the purpose of this paper.  Both have idempotents and it is through the idempotents that we can see how the non-Boolean structure arises.

In our approach, the idempotent is the analogue of the projection operator in the standard approach and it is the projection operators that lead to a propositional logic, exploited by Birkhoff and von Neumann \cite{gbvjn}.  They were the first to suggest that quantum theory should be regarded as a new `logic', quantum logic, which has now been developed into a formal structure \cite{jbmm77}. However this has not been very popular amongst physicists and chemists because it has not led to any new ways of thinking about quantum phenomena.  Nevertheless it is necessary to understand how this structure arises before we can establish a different point of view.  First it is necessary to know that the structure of any von Neumann algebra is determined by its idempotents, or, if you like, its projection operators.   Both idempotents and projection operators  have two eigenvalues, $(0,1)$, so we can regard them as propositions giving us a truth value.  Because of the non-commutativity of idempotents, quantum theory gives rise to a non-Boolean propositional calculus \cite{jj68}. 

The problem with this interpretation is that it appears to become epistemological, that is it has to do with questions that we ask of the physical system.  As a physicist,  I am interested in the ontology underlying quantum phenomena.  I will follow Eddington \cite{ae58} and introduce a {\em structural concept of existence} rather than relying on some metaphysical concept of a particle.  Existence manifests itself in two ways--it either exists or it doesn't.  Thus let us associate existence with an idempotent.  This seems an eminently good notion that has very general applicability.  For example if I consider who I am, where is the real me?  My mind is constantly in turmoil, my body, my cells and even my bones are actively changing their constituents.  I inhale and exhale, etc.  I am in constant change, yet it is still me.  I am an idempotent, constantly changing into myself.  Yes, there are small changes over time, so that the idempotent can change over time, but time scales become crucial particularly at the sub-atomic scales, where `particles' exist for very short times.  The notion of relative stability becomes primary at this level and it is here that processes exhibit fleeting existences.

If we follow this route, then non-commuting structures throw up very interesting consequences.  Idempotents do not necessarily commute which implies that when some processes are manifest, others can be completely undefined, so that we cannot even ask if they exist or not.  But there is more.  Since there exist inner automorphisms in the algebras, we can relate non-commuting idempotents so that, for example we can write $\epsilon'=A\epsilon A^{-1}$.  In order to see what this means, let us consider a matrix representation of the structure\footnote{All type I von Neumann algebras have matrix representations.}. Then
\begin{eqnarray*}
\epsilon'_{jj}=\sum_{k}A_{jk}\epsilon_{kk}A_{kj}^{-1}
\end{eqnarray*}
Thus we see that each transformed $\epsilon_{jj}'$ contains contributions from {\em all} the idempotents in the set $\{\epsilon\}$.  We have called this the {\em exploding transformation}\footnote{This is the structure used in the Huygens construction and hence the Feynman path integral method \cite{fh12}}.  What this transformation implies is that the idempotent $\epsilon_{jj}'$ contains contributions from all the idempotents $\{\epsilon\}$.  In terms of existence, this means that when we make a transformation, it is not that the existing entity signified by one idempotent `vanishes into thin air', as it were, but that it contributes and is active in the new idempotent.

If we only think in terms of classical materialism, then this idea makes little sense, but if we think of process, then it implies that everything is an undivided whole but within that totality there exist invariants, the invariants that give rise to quasi-local, semi-stable structures to which we give the name `particle'.  These semi-stable structures can come together and form even more stable structures through their mutual interaction.  It is out of these stable structures that the classical world emerges. After all, as has already been pointed out, quantum mechanics was introduced to explain the stability of the macroscopic world.

Thus the individual `particle' exists only in the background of the total process.  As Primas puts it ``The environment must never be left out of consideration" \cite{hp75}.  However it is actually stronger than that.  Without the background there would be no invariant, there would be no particle.  This is totally different from the classical view where we assume the particle exists {\em a priori} as an autonomous preexistent object.

\section{Example of Non-commutative Probability in Quantum Mechanics}

I want to continue by exploring the appearance of non-commutative probability in the von Neumann/Moyal algebraic approach.  Let us follow Moyal and Feynman \cite{rf87} and regarding $f_\rho(\alpha, \beta)$ as a probability measure even though it can take negative values\footnote{See Bartlet \cite{mb45} for a discussion of this point.}.  Here we will identify $\alpha =x, \beta = p$ and take $\rho$ to be the density matrix for a system in a pure state $\psi(x,t)$ so that
\begin{eqnarray*}
f_\psi(x,p)=\frac{1}{2\pi}\int\psi^*(x-\tau/2)e^{-ip\tau}\psi(x+\tau/2)d\tau.
\end{eqnarray*}
This will be recognised as the Wigner function which is easily obtained from the two-point density matrix \cite{dbbh81}.  In this case the parameters $(x,p)$ are the mean coordinates of what de Gosson calls a {\em quantum blob} \cite{mdg12}.

 Since the probability measure depends upon two variables, we can ask for the conditional expectation of, say,  the momentum at a point $x$.  Moyal \cite{jm49} shows this is
\begin{eqnarray*}
\rho(x)\bar p=\int pf_\psi(x,p)=\left(\frac{1}{2i}\right)[(\partial_{x_1}-\partial_{x_2})\psi(x_1)\psi(x_2)]_{x_1=x_2=x}
\end{eqnarray*}
If we write $\psi=Re^{iS}$ we find
\begin{eqnarray*}
\bar p(x, t)=\nabla S(x,t)
\end{eqnarray*}
which is just the so called ``guidance condition" used in the Bohm approach to quantum mechanics \cite{dbbh93} only in this context it is not guiding anything.  Here it is simply a conditional expectation value of the momentum.  

Moyal also obtains an equation for the transport of this momentum.  Starting from Heisenberg's equation of motion, Moyal finds
\begin{eqnarray*}
\partial_t(\rho\bar p_k)+\sum_i\partial_{x_i}(\rho p_k\partial_{x_i}H)+\rho\partial_{x_k}H=0
\end{eqnarray*}
Once again if we write $\psi=Re^{iS}$, we find
\begin{eqnarray*}
\frac{\partial}{\partial x_k}\left[\frac{\partial S}{\partial t}+H-\frac{\nabla^2\rho}{8m\rho}\right]=0
\end{eqnarray*}
Or
\begin{eqnarray*}
\frac{\partial S}{\partial t}+H-\frac{\nabla^2\rho}{8m\rho}=\frac{\partial S}{\partial t}+\frac{1}{2m}(\nabla S)^2-\frac{1}{2m}\frac{\nabla^2R}{R}=0
\end{eqnarray*}
where we have chosen the constant of integration to be zero.  This equation is just the quantum Hamilton-Jacobi equation exploited in the Bohm approach.  This equation is simply a conservation of energy equation where $\nabla^2R/2mR$, is the quantum potential which is regarded as a new quality of energy.  One can show that the quantum potential is related to the $T_{jj}$ component of the energy-momentum tensor of the Schr\"{o}dinger field \cite{tt53}.

By integrating $\int \nabla S dt$,  a set of stream lines can be calculated as first shown in Philippidis, Dewdney and Hiley \cite{cpcdbh79} for the classic two-slit interference pattern.  Other typical quantum situations are discussed in Bohm and Hiley \cite{dbbh93}, Holland \cite{ph93} and Wyatt \cite{rw05}.   If we assume, with Bohm that the particle actually possesses this momentum, then we have the possibility of understanding the behaviour of individual particles.  What we have done here is to represent the particle in a phase space spanned by the co-ordinates $(x, \bar p)$ and have calculated an ensemble of `trajectories' along which the particle could travel.

Alternately we could examine the conditional expectation value of $x$ which is given by 
\begin{eqnarray*}
\rho\bar x=\int xf_\phi(x,p)dx=\frac{1}{2i}[(\partial_{p_1}-\partial_{p_2})\phi^*(p_1)\phi(p_2)]_{p_1=p_2=p}
\end{eqnarray*}
where $\phi(p)$ is the Fourier transformation of the wave function $\psi(x)$.  Again if we write $\phi(p)=R(p)e^{iS(p)}$, we find 
\begin{eqnarray*}
\bar x(p)=-\frac{\partial S_p}{\partial p}.
\end{eqnarray*}
This relation replaces the so called guidance condition, but note here there is no way that this expression can be regarded as ``guiding" anything.  Nevertheless  we  construct a new phase space, this time with co-ordinates $( \bar x,p)$.  One can also calculate stream lines in this space \cite{mbbh00} and one finds that the streamlines are different in the two cases.  The question is then how do we reconcile these differences?

\subsection{Shadow Manifolds.}

In order to explain the appearance of these two phase spaces, we must recall the Gelfand-Naimark construction \cite{jv06}.  This construction requires us to think of the evolution of material processes an entirely new way.  Rather than starting from an {\em a priori} given space-time with its preassigned topological and metrical properties upon which the algebraic structure that describes the evolution of the material process, we start from the algebraic structure and then abstract the properties of the underlying manifold.  If the dynamical algebraic structure is commutative then the Gelfand-Naimark theorem tells us that there is a unique underlying manifold whose topological and metrical structure are determined by the dynamical algebra,
In this case the points of the space are maximal two-sided ideals of the algebra so that the points of the space are part of the algebra itself.   Thus the space of points are not separate entities but are part of the whole structure

While this works for a commutative structure, one finds no unique underlying manifold if the algebra is non-commutative.  The best one can do is to abstract out a set of {\em shadow manifolds}.  The two phase spaces that we constructed above are examples of these shadow manifolds.  This can be taken to be a rigorous mathematical statement of Bohr's principle of complementarity.  It does not need to be considered as  ``wave-particle" duality, a notion, although popular, makes very little sense when carefully examined.   In our view the non-commutative structure is the ontological structure which captures more clearly the notion of wholeness that Bohr felt was an essential feature of quantum phenomena.  But if the structure is non-Boolean and you are trying to explain it in terms of a Boolean logic, then
the two alternative structures arise merely because we are trying to project the process into an inappropriate descriptive form.  

At this stage the idea is being discussed in terms of just two shadow spaces.  However it can be shown that there could be many shadow spaces.  For example the two spaces arise from the Fourier transform but mathematically we could use the fractional Fourier transformation, then we can obtain a family of shadow manifolds as shown in Brown \cite{mb04}.

\begin{figure}[htbp] %  figure placement: here, top, bottom, or page
   \centering
   \includegraphics[width=1in]{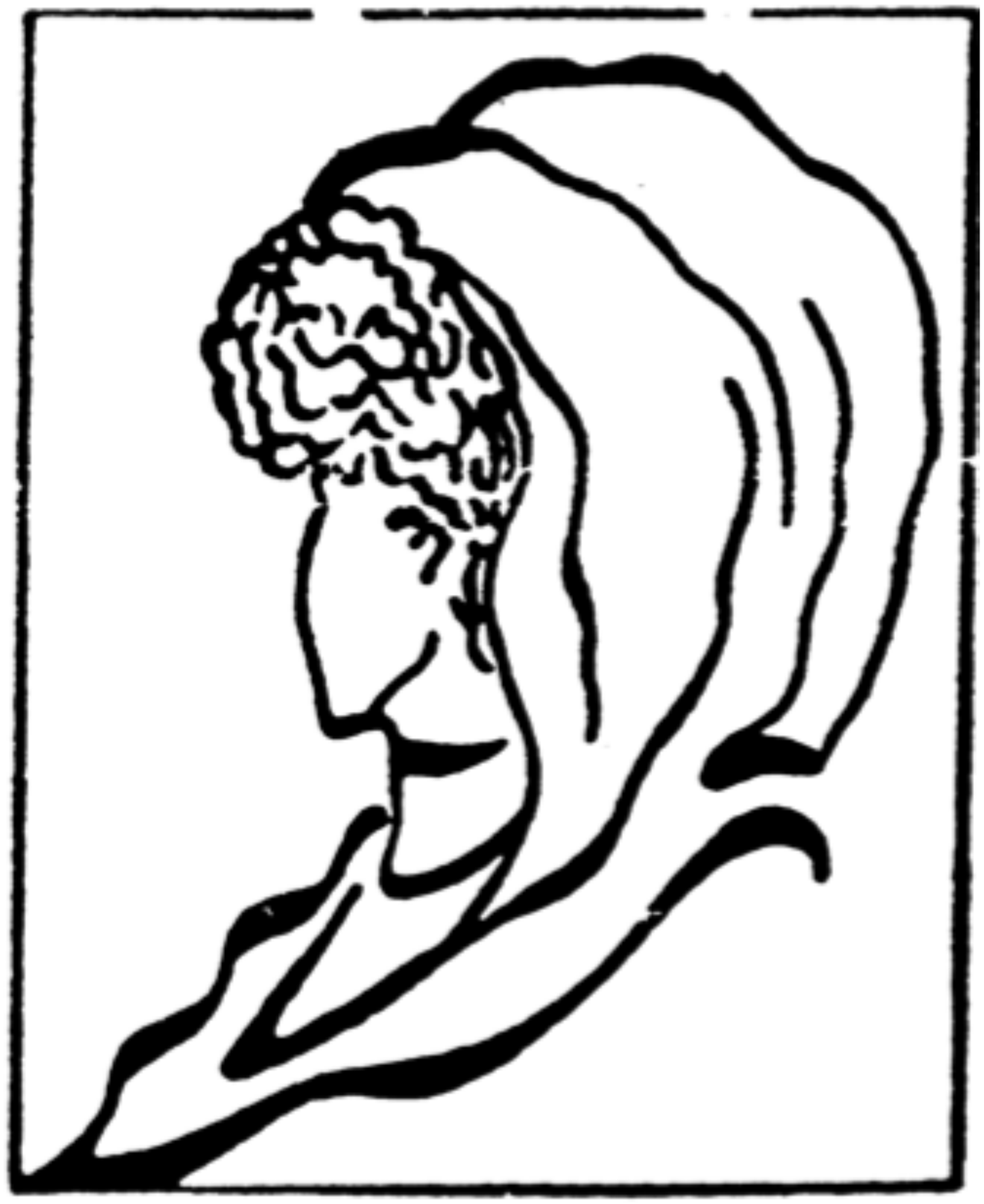} 
   \caption{Old Lady-Young Lady.}
   \label{fig:1}
\end{figure}

If one wants to consider this in philosophical terms then the non-commutative algebra is essentially a description of the implicate order, while the shadow manifolds are merely the explicate orders \cite{db80}.  The way I have tried to get this view across is to recall the gestalt effect revealed in pattern or drawing in which we can see two alternative figures.  A typical well known example is the old lady-young lady image shown in Figure \ref{fig:1}.

Here it is the observer that is trying to find some meaning in the drawing.  Of course that does not mean that we descend into some form of subjectivism.  The drawing is real, we are simply trying to make sense of it.

It is interesting to note that Primas \cite{hp75}, by recognising the holistic nature of quantum phenomena, also argues that phenomena or patterns have no {\em a priori} meaning.  We provide the meaning in the same sense that we provide the meaning to Figure \ref{fig:1}.  Naturally the meaning is subjective but the pattern or the phenomena is not.  That is real; that is ontological.  Primas goes on to argue that pattern recognition is a map from the non-Boolean world into a Boolean description.  This is exactly what Bohm \cite{db80} was getting at with his implicate-explicate order.  The explicate orders are Boolean accounts that emerge from the non-Boolean world of quantum phenomena. 

\section{ Conclusion.}

We have taken the algebraic description of quantum phenomena to illustrate how non-commutative probability theory applies to the material world.  But the general structure of the idea has a much wider application and holds even in the world of thought.  We all experience the struggle to explicate our thoughts and feelings!  We are forced into explanations that are precise, are Boolean, but our thought is not Boolean.  We cannot give {\em one} view of reality, not because we, as humans, are limited in our in our abilities or that we are ``clumsy" in the laboratory disturbing everything we try to explore.  We are limited because nature is holistic and does not allow a reductionist view of nature except in a somewhat limited domain, limited but vital for our immediate survival.  However the deeper lessons that we learn about material reality, hold even more so when it comes the mental world.

It is not that the mental world is separate from the physical world.  They are both aspects of the same underlying structure process.  I don't have the time here to discuss this further but this aspect has been eloquently argued by Bohm \cite{db90}, an argument that I will not repeat here.  I hope that this paper will begin to redress those who argue that the lessons of quantum theory have nothing to teach us about these much deeper questions, particularly those addressing the relation between mind and matter.  By bringing out the deeper structure of the ideas, we do not waste the opportunity by being trapped in arguments that claim the brain is too hot and too wet for these ideas to be relevant.

 %References 1

\end{document}